\documentclass[hypertex,
letterpaper,twocolumn,showpacs,preprintnumbers,amsmath,amssymb]{revtex4}

\usepackage{graphicx}
\usepackage{dcolumn}
\usepackage{bm}
\usepackage{amsmath}
\newcommand{\fr}[2]{\textstyle{\frac{#1}{#2}}}
\newcommand{\tin}[1]{{\mbox{\tiny $ #1 $}}}
\newcommand{\mui}{\mu_{\mbox{\tiny I}}}
\newcommand{\llang}{\langle\!\langle}
\newcommand{\rrang}{\rangle\!\rangle}

\usepackage[colorlinks=true,citecolor=blue]{hyperref}

\begin{document}

\preprint{}

\title{Two-flavor condensates in chiral dynamics: temperature and isospin density effects. }

\author{M. Loewe}
\author{C. Villavicencio}%
\affiliation{%
 Facultad de F\'{\i}sica,
 Pontificia Universidad Cat\'olica de Chile,
 Casilla 306, Santiago 22, Chile}%

\begin{abstract}
Isospin density and thermal corrections  for several condensates
are discussed, at the one-loop level, in the frame of chiral
dynamics with pionic degrees of freedom. The evolution of such
objects give an additional insight into the condensed-pion phase
transition, that occurs basically when $|\mui|>m_\pi$, being
$|\mui|$ the isospin chemical potential. Calculations are done in
both phases, showing a good agreement with lattice results for
such condensates.
\end{abstract}


\maketitle

This paper is an extension of our previous analysis of pion
dynamics, according to chiral perturbation theory, in the presence
of isospin chemical potential and temperature. In the first
article \cite{Loewe:2002tw} we discussed the evolution of the 
masses and decay
constants from the perspective of the first phase
($|\mui|<m_\pi$). In the second paper \cite{Loewe:2004mu}, we
proposed a scheme of calculation in the second phase
 for the pion masses ($|\mui|>m_\pi$). The method allowed us to
explore the regions where the chemical potential is close to the
phase transition point ($|\mui|\gtrsim m_\pi$) and also where
$|\mui|\gg m_\pi$. The validity of this approach is restricted to
values of $\mui$ less than the $\eta$ or $\rho$ masses.

Here we would like to address the thermal and density behavior of
several condensates that can be buit in this frame, as well as
the validity of the Gell-Mann--Oakes--Renner (GMOR) relation,
completing in this way the discussion of the pion properties. We
compare our results with those obtained through lattice
measurements.

This analysis is relevant since some of these condensates are
sharp signals for the occurrence of the phase transition, i.e.,
phenomenological order parameters. Natural scenarios where this
dynamics can play a role are in the core of neutron stars (especially
during the cooling period), possible asymmetries in the pion
multiplicity in the central rapidity region at RHIC or ALICE, etc.

As in the previous articles, we introduce the chemical potential
following \cite{Kogut:1999iv,Kogut:2000ek}. Even though, these
articles deal with QCD with two colors rather than QCD with three
colors, it is clear that both problems are intimately related. The
introduction of in-medium processes via isospin chemical potential
has been studied at zero temperature
\cite{Son:2000xc,Kogut:2001id} in both phases ($|\mui|\lessgtr m_\pi$) at tree
level.

Different approaches, such as Lattice QCD
\cite{Kogut:2002tm,Kogut:2002zg,Kogut:2001if,Kogut:2002cm}, Ladder
QCD \cite{Barducci:2003un} and Nambu-Jona-Lasinio model based
analysis
\cite{Toublan:2003tt,Frank:2003ve,Barducci:2004tt,Barducci:2004nc,He:2005sp}
have confirmed the appearance of an interesting and non-trivial
phase structure as function of temperature and chemical
potentials, in particular isospin, chemical potential.

\section{Chiral Lagrangian}\label{chi_lag}

In the low-energy description where only pion degrees of freedom
are relevant, the most general chiral invariant lagrangian at the
second order, ${\cal O}(p^2)$, according to an expansion in powers
of the external momentum is given by
\begin{eqnarray}
    {\cal L}_2  &=& \frac{f^2}{4}Tr\left[(D_\mu U)^\dag D^\mu U+U^\dag \chi
+ \chi^\dag U\right]
\end{eqnarray}
with
\begin{eqnarray}
 D_\mu U &=& \partial_\mu U-i[v_\mu, U]-i\{ a_\mu, U\},\\
 \chi &=& 2B(s+ip),\\
 U &=& \bar U^\tin{1/2}(e^{i\pi^a\tau^a/f})\bar U^\tin{1/2},
\end{eqnarray}
$s=s^0(x)+s^a(x)\tau^a$, $p=p^0(x)+p^a(x)\tau^a$,
$v_\mu=\frac{1}{2}v_\mu^a(x)\tau^a$ and
$a_\mu=\frac{1}{2}a_\mu^a(x)\tau^a$ being the scalar,
pseudo-scalar, vector
and axial-vector external sources, respectively, where $\tau^a$
are the Pauli matrices. $\bar U$ is the vacuum expectation value
of the field $U$. $M=diag(m_u,m_d)$ is the quark mass matrix and
$B$ in the previous equation is an arbitrary constant which will
be fixed when the mass is identified setting $(m_{u} + m_{d})B
=m^{2}$, where $m$ denotes the bare (tree-level) pion mass. We
will use $m_\pi$ to denote the pion masses after renormalization.

 The most general ${\cal O}(p^4)$ chiral lagrangian has the form

\begin{eqnarray}
{\cal L}_4 &=&  \fr{1}{4}l_1\left( Tr\left[(D_\mu U)^\dag D^\mu
           U\right]\right)^2  \nonumber\\
 && +  \fr{1}{4}l_2Tr\left[(D_\mu U)^\dag D_\nu
           U\right]Tr\left[(D^\mu U)^\dag D^\nu U\right]\nonumber\\
 &&+  \fr{1}{16}(l_3+l_4)\left(Tr\left[\chi U^\dag
           +U\chi^\dag\right]\right)^2\nonumber\\
 &&+  \fr{1}{8}l_4Tr\left[(D_\mu U)^\dag D^\mu U\right]Tr\left[\chi
           U^\dag + U\chi^\dag    \right]\nonumber\\
 &&+  l_5Tr\left[\big(L_{\mu\nu}U +UR_{\mu\nu}\big)\big(U^\dag
           L^{\mu\nu}+R^{\mu\nu}U^\dag\big)\right]\nonumber\\
 &&  + l_6Tr\left[iL_{\mu\nu}D^\mu U(D^\nu U)^\dag +iR_{\mu\nu}(D^\mu
           U)^\dag D^\nu U\right]\nonumber\\
 && -\fr{1}{16}l_7\left(Tr\left[\chi
           U^\dag-U\chi^\dag\right]\right)^2
 +\fr{1}{4} (\tilde{h}_1+\tilde{h}_3) Tr[\chi^\dag\chi]
\nonumber \\
 &&
 -\fr{1}{8}(\tilde{h}_1-\tilde{h}_3)Tr\left[\left(\chi-Tr[
\chi]
\right)^2+\left(\chi^\dag-Tr[\chi^\dag]\right)^2\right]
\nonumber\\
&&
-2\tilde{h}_2Tr\left[L_{\mu\nu}L^{\mu\nu}+R_{\mu\nu}
R^{\mu\nu}\right],
\end{eqnarray}

The $l_i$ are the original  Gasser and Leutwyler coupling
constants \cite{Gasser:1983yg} for an $SU(2)$ Lagrangian, and the
$\tilde h_i$ constants are couplings to pure external fields, and
their values are model dependent. They have to be determined
experimentally and are also tabulated in several articles and
books. This Lagrangian includes the chemical potentials in the
covariant derivatives and in the expectation value of the $\bar U$
matrix. The constants $l_i$ include divergent corrections which
allow to cancel the divergences from loops corrections constructed
from the ${\cal L}_2$ Lagrangian

\begin{equation}
l_i(\Lambda)=\frac{\gamma_i}{32\pi^2}\left[\bar l_i-\lambda
  +\ln\left(\frac{m^2}{\Lambda^2}\right)\right]
\end{equation}
with the $\overline{ms}$ pole
\begin{equation}
\lambda=\frac{2}{4-d}+\ln 4\pi +\Gamma^\prime (1)+1.
\end{equation}
The $\tilde h_i$ constants do not include divergent terms.

For $|\mu_\tin{I}|>m$ there is a symmetry breaking. The vacuum
expectation value that minimizes the potential, calculated in
\cite{Kogut:2001id},  is
\begin{eqnarray}
\bar U
 &=&
  \fr{m^2}{\mu_\tin{I}^2}
  +i[\tau_1\cos\phi+\tau_2\sin\phi]
   \sqrt{1-\fr{m^4}{\mu_\tin{I}^ 4}}\nonumber\\
 &\equiv&
  c+i\tilde\tau_1 s,
 \label{U}
\end{eqnarray}
where $c\equiv \frac{m^2}{\mu_\tin{I}^2}$,
$s\equiv\sqrt{1-\fr{m^4}{\mu_\tin{I}^ 4}}$. From now on, we will
use a tilde to refer to any vector rotated in a $\phi$ angle
\begin{eqnarray}
\tilde v_1 &=& v_1\cos\phi+v_2\sin\phi\nonumber\\
\tilde v_2 &=& -v_1\sin\phi+v_2\cos\phi.
\end{eqnarray}

For our purposes it is enough to keep the ${\cal L}_{2,2}$ terms
of the lagrangian to construct the Dolan-Jackiw propagators (DJp)
which are the same as the ones we used in the previous articles, both in the
first and in the second phase, respectively
\cite{Loewe:2002tw,Loewe:2004mu}

In both phases the $\pi_0$ propagator will be the same, because it
is always diagonal
\begin{equation}
D_{DJ}(p)_{00} = \frac{i}{p^2 -m^2 +i\epsilon} +2\pi
  n_B(|p_0|)\delta (p^2 -m^2)
\end{equation}
In the first phase, where the charged pions do not mix, the
thermal propagators are
\begin{multline}
D_{DJ}(p)_{+-} = \frac{i}{(p +\mui u)^2 -m^2 +i\epsilon}\\
  +2\pi n_B(|p_0|)\delta\big( (p +\mui u)^2 -m^2\big)
\label{D+-}
\end{multline}
and $D_{DJ}(p)_{-+} = D_{DJ}(-p)_{+-}$.Those propagators refer to
the fields  $\pi_+$ and $\pi_-$. In the second phase, where the
charged pion fields become mixed, the propagators have a
non-trivial matrix structure (see \cite{Loewe:2004mu}). By scaling
all the parameters and variables with $|\mui|$, it is possible to
make an expansion in terms of an appropriate smallness parameter
in powers of $s^n$ when $|\mui|\gtrsim m$ and $c^n$ when
$|\mui|\gg m$, where $s$ and $c$ are defined in eq. (\ref{U}). A
similar expansion criteria was proposed earlier by Splittorff,
Toublan and Verbaarschot
\cite{Splittorff:2001fy,Splittorff:2002xn}. In this case we will
use the lowest order term in the expansion and a negative
isospin chemical potential $\mui=-|\mui|$, as in the case of
neutron stars.

The DJp in the region where $|\mui|\gtrsim m$ are the same as in
eq. (\ref{D+-}) plus corrections of ${\cal O}(s^2)$ referring to
the fields
 $\tilde{\pi}_+$ and $\tilde{\pi}_-$. The other propagators 
 $D_{DJ}(p)_{++} = D_{DJ}(p)_{--}$ are of order $s^2$.

In the case where $|\mui|\gg m$, it is better to work with the
fields $\tilde{\pi}_1$ and $\tilde{\pi}_2$ instead of
$\tilde\pi_\pm$. The DJp for the $\tilde\pi_1$ field is the same
as the the $\pi_0$ propagator plus higher corrections:
$D_{DJ}(p)_{11}=D_{DJ}(p)_{00}+\mui^{-2}{\cal O}(c^2)$. For the
$\tilde\pi_2$ propagator we have
\begin{equation}
D_{DJ}(p)_{22} = \frac{i}{p^2 +i\epsilon}
  +2\pi n_B(|p_0|)\delta (p ^2) +\mui^{-2}{\cal O}(c^2),
\end{equation}

The last propagator we will introduce is the mixed
$\tilde\pi_1\tilde\pi_2$ propagator, which will be used in the
loop calculation for the axial-charge density condensate
\begin{eqnarray}
D_{DJ}(p)_{12} &=& -2ic\frac{p_0}{|\mui|} \bigg\{\frac{i}{p^2-\mui^2+i\epsilon}
-\frac{i}{p^2+i\epsilon}\nonumber\\
&&+2\pi n_B(|p_0|)\big[\delta (p^2-\mui^2)-\delta (p^2)\big]\bigg\} \nonumber\\
&&+\mui^{-2}{\cal O}(c^3).
\end{eqnarray}
As in the case of the $\pi_\pm$ propagators, which are
anti-diagonal in the propagator matrix, we have
$D_{DJ}(p)_{21}=D_{DJ}(-p)_{12}$.
The other possible propagators are zero in both phases:
$D_{DJ}(p)_{3,a\neq 3}\equiv 0$

\section{Condensates}

The different condensates can be constructed taking  appropriate
functional derivatives with respect to the external sources in the
extended chiral lagrangian. In this way we will consider the
following  currents
\begin{align}
V_\mu^a = \frac{\delta S}{\delta v_\mu^a}
&= {\fr{1}{2}}\bar q\gamma_\mu\tau^a q,
& A_\mu^a = \frac{\delta S_\chi}{\delta a_\mu^a}
&= {\fr{1}{2}}\bar q\gamma_\mu\gamma_5\tau^a q,\nonumber\\
J_p^0 = \frac{\delta S_\chi}{\delta p^0}
&= \bar q\gamma_5q,
& J_p^a = \frac{\delta S_\chi}{\delta p^a} &=\bar q\gamma_5\tau^aq,\nonumber\\
-J_s^0 =  \frac{\delta S_\chi}{\delta s^0} &= -\bar qq, &
-J_s^a =
\frac{\delta S_\chi}{\delta s^a} &= -\bar q\tau^aq,
\end{align}
where $S_\chi(s,p,v,a)$ is the action of our extended chiral
lagrangian. After computing the different currents, the
introduction of isospin chemical potential for massive states is
achieved by taking the following values of the external sources:
$s=M$, $v_\mu=\frac{1}{2}\mui\tau_3u_\mu$ and $a_\mu=p=0$, where
$M$ is the quark-mass matrix and $u_\mu$ is the fourth velocity.

The vacuum expectation values of these objects are precisely the
condensates we would like to analyze. Only some of these
condensates will behave in a non-trivial  way with respect to
$\mui$ and $T$. Here we will compute the one-loop corrections to
these condensates in the two phases.

The resulting currents are sorted in powers of momentum:
$J=J_{(1)}+J_{(3)}+\dots$. Expanding then in powers of fields, the
effective current will be of the form $J_{\tin{eff}}=
J_{(1,0)}+J_{(1,1)}+J_{(1,2)}+J_{(3,0)}$. The notation $J_{n,m}$
refers to the power counting of $P^n$ (momentum powers) and
$\pi^m$ (field powers). In this case the term $J_{(1,1)}$ is not necessary
since the vaccum expectation value of a single pion field vanishes: $\llang
0|\pi^a|0\rrang\equiv 0$. We refer $|0\rrang$ to the thermal vacuum or
``populated" vacuum.

By considering $S_2$ and $S_4$, it is not difficult to realize
that $\llang 0|J_p^0|0\rrang\equiv 0$ in both phases. For the case
of $\llang 0|J_s^a|0\rrang $ we find, as expected, the result by
Gasser and Leutwyler \cite{Gasser:1983yg}
\begin{equation}
\llang 0|J_s^a|0\rrang =\llang 0|\bar uu-\bar dd| 0\rrang
\delta^{a3}= 4m^2B\epsilon_{ud}\tilde h_3\delta^{a3},
\end{equation}
which depends only on the ${\cal L}_4$ coefficients. So, at the
tree level, this quantity vanishes. $\epsilon_{ud}\equiv
(m_u-m_d)/(m_u+m_d)$ is a small quantity. The important thing is
that this condensate does not depend on temperature and/or isospin
chemical potential, and therefore it does not play the role of an
order parameter for the phase transition.

\section{First phase: $|\mui|<m$}

In this phase the only non-vanishing condensates are the chiral
condensate and the isospin number density. The other condensates
(pion condensate and axial charge density) vanish due to the
trivial structure of the vacuum in this phase where parity is
conserved.

\subsection{Chiral condensate.}

The chiral condensate is the natural order parameter associated
the chiral symmetry breaking. It corresponds to the non-vanishing
vacuum expectation value of the scalar current
\begin{equation}
\llang\sigma\rrang\equiv\llang 0|\bar qq|0\rrang = \llang
0|J_s^0|0\rrang .
\end{equation}

The components of the effective current relevant for radiative
calculations are
\begin{eqnarray}
 J_{s(1,0)}&=&-2Bf^2,\nonumber\\
 J_{s(1,2)}&=&B[\pi_0^2+2|\pi|^2],\nonumber\\
 J_{s(3,0)}&=&-4Bm^2(l_3+l_4+\tilde h_1),
\end{eqnarray}
where $|\pi|^2\equiv \pi_+\pi_-$. Using the fact that
\begin{equation}
\llang 0|\pi_a(x)\pi_b(x)|0\rrang=\llang
0|T\pi_a(x)\pi_b(y)|0\rrang_{y=x},
\end{equation}
the corrected chiral condensate turns out to be
\begin{equation}
\llang\sigma\rrang =-2Bf^2\big\{1 -\alpha\big[\fr{1}{2}\bar
l_3-2\bar l_4-32\pi^2\tilde h_1+2I_0+4I\big]\big\},
\label{phase1.qq(T,mu)}
\end{equation}
where the divergences were removed by the ${\cal L}_4$
counterterms and $I$, $I_0$ are integral functions of $T$ and
$\mui$, which are tabulated in the appendix.

The constant $\tilde h_1$, as was said before, is model dependent.
From the G-MOR relation at finite temperature and zero chemical
potential
\begin{equation}
m_\pi^2f_\pi^2=-\fr{1}{2}(m_u+m_d)\llang\sigma\rrang,
\end{equation}
and using the results for the evolution of the chiral condensate
in the previous equation as well as the $m_\pi(T)$ and $f_\pi(T)$
given in \cite{Loewe:2002tw} (and also in other papers) we can see
that the G-MOR relation remains valid if we neglect the term
$\epsilon_{ud}^2l_7\sim 0$ that appears in the $\pi_0$ mass
corrections, and take $\tilde h_1=0$.
  If we take into account higher corrections to the G-MOR relation at
finite temperature, the $\tilde h_1$ can be associated with the
continuum threshold that appears in the sum-rules approach
\cite{Dominguez:1996kf}.

\begin{figure}
\includegraphics[scale=.8]{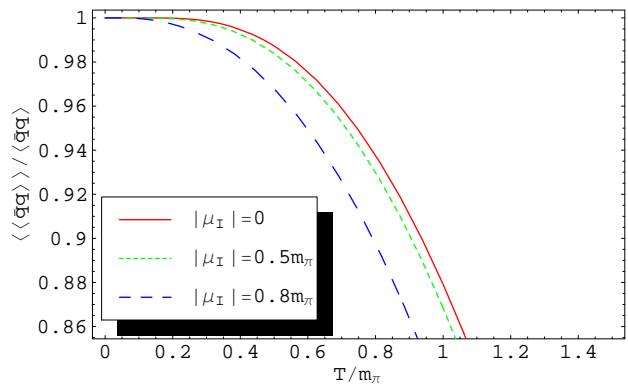}
\caption{ Quark condensate as a function of the temperature and
the isospin chemical potential} \label{phase1.figure.qq}
\end{figure}

Figure \ref{phase1.figure.qq} shows the chiral condensate at
finite temperature and isospin chemical potential. It is
interesting to remark that the chiral condensate vanishes also,
almost linearly with the baryonic density \cite{Peng:2003nq}.

The introduction of  isospin chemical potential breaks the
degeneracy of the thermal evolution of the pion masses and decay
constants \cite{Loewe:2002tw}.  This suggests considering the
G-MOR relation in terms of the average value
 for the masses and decay constants. In fact, using our results
 from the above quoted paper we find that
\begin{equation}
\overline{m}_\pi^2\overline{f}_\pi^2=-\fr{1}{2}(m_u+m_d)\llang\sigma\rrang.
\end{equation}
where
\begin{eqnarray}
\overline{m} _\pi &\equiv& \fr{1}{3}\left(m_{\pi^+}
+m_{\pi^-}+m_{\pi^0}\right),\\
\overline{f}_\pi&\equiv&\fr{1}{3}\left(f_{\pi^+}
+f_{\pi^-}+f_{\pi^0}\right).
\end{eqnarray}

With this prescription for the G-MOR relation, we can consider
 $\tilde h_1 =\frac{1}{3}\epsilon_{ud}^2l_7 \sim 0$.

\subsection{Isospin number density.}

The isospin number density condensate gives us information about
the baryon number difference between $u$ and $d$ quarks in the
thermal or populated vacuum. The isospin number density is
defined as the 0-Lorentz component and third-isospin component of
the vector current
\begin{equation}
\llang n_I\rrang\equiv \llang 0|\fr{1}{2}q^\dag\tau^3q|0\rrang =\llang
0|V_0^3|0\rrang.
\end{equation}

\begin{figure}
\includegraphics[scale=.77]{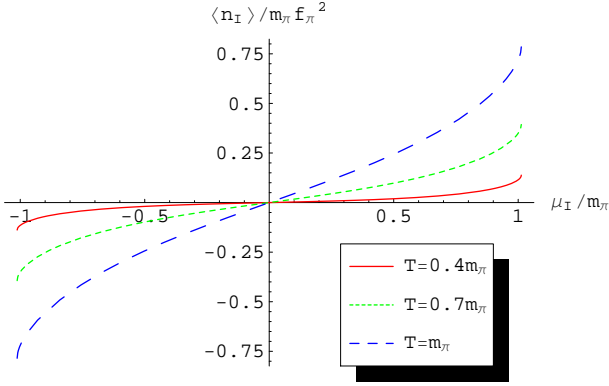}
\caption{Isospin-number density condensate as a function of the
isospin chemical potential for different values of the
temperature.} \label{fig.niphase1}
\end{figure}

We need to calculate the expectation value of the vector current
in the vacuum. We will see that the only non-vanishing component
of the vector current in the vacuum is the isospin number density.
In fact, by considering the expansion of the vector current and
using the fact that $\llang 0|\pi_a|0\rrang =0$ and $ \llang
0|\pi_{a\neq 3}\pi_3|0\rrang =0$
 we find
in the first phase case
\begin{multline}
 \llang 0|\bm{V}_{(1,2)\mu}|0\rrang = \llang
0|-i(\pi_\tin{+}\partial_\mu\pi_\tin{-}
-\pi_\tin{-}\partial_\mu\pi_\tin{+})\\
 +2\mui|\pi|^2u_\mu|0\rrang{\bf e}_3.
 \end{multline}
This means that the isospin-number density depends on temperature
only through thermal
 insertions in loop corrections, and therefore,
  vanishes for $T=0$ or $|\mui|=0$.
 Using the fact that
 \begin{equation}
 \llang 0|\pi^a(x)\partial\pi^b(x)|0\rrang
 =\Big[\partial_y\llang 0|T\pi^a(x)\pi^b(y)|0\rrang\Big]_{y=x},
 \end{equation}
we find that the vacuum expectation value of the vector current is oriented in
the ${\bf e}_3$ direction and proportional to the fourth velocity: $\llang
0|\bm{V}_{\mu}|0\rrang =\llang n_I\rrang u_\mu{\bf e}_3$. The isospin number
density condensate is then.
\begin{equation}
\llang n_I\rrang = 8mf^2\alpha\epsilon(\mui)J,
\end{equation}
where $J$ is a function of $T,\mui$  defined in the Appendix. A
finite value of this condensate requires a finite value for both,
$\mui$ and $T$, as can be seen in fig. \ref{fig.niphase1}

\section{Second phase: $|\mui|\gtrsim m$}

In this phase, due to the non trivial vacuum expectation value of
the $U$ fields mentioned in eq. (\ref{U}), a condensation of
pions, in this case of negative pion, occurs, giving rise to a
superfluid phase. The apparition of two new condensates, the pion
condensate and the axial-charge density condensate will break
parity.

As we mentioned before in section \ref{chi_lag}, we can expand the
loop corrections in this region as a series of powers of
$s^2=1-\frac{m^4}{\mui^4}$. The condensates will have the shape
\begin{equation}
\llang J\rrang =\langle J\rangle \bigg\{\eta
(s)+\alpha'\sum_{n=0}\sigma_n(T/|\mui|)s^{2n}\bigg\} \label{cond.s}
\end{equation}

\subsection{Chiral condensate}

Following the same procedure we used in the case of the first
phase, the non-vanishing components of the chiral condensate at
order $s^0$ are
\begin{eqnarray}
J_{s(1,0)}&=&-2Bf^2c,\\
J_{s(1,2)} &=& Bc \big[\pi_0^2+2|\tilde\pi|^2\big],\\
J_{s(3,0)} &=&-4B\mui^2c[l_3+l_4+\tilde h_1+{\cal O}(s^2)].
\end{eqnarray}
The term $c$ must be kept because it is a global constant.

The resulting chiral condensate at finite temperature and isospin
chemical potential is then
\begin{equation}
\llang\sigma\rrang = -2Bf^2c\big\{1+\alpha'\big[-\fr{1}{2}\bar
l_3+2\bar l_4-2I'_0-4I'\big]\big\}.
\end{equation}

\begin{figure}
\includegraphics[scale=.75]{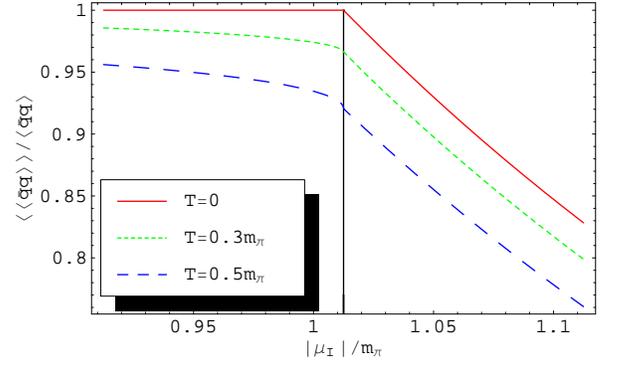}
\caption{ Chiral condensate as a function of the
isospin chemical potential for different values of the
temperature. The vertical line denotes the transition point
between the two phases.} \label{figure.qqs}
\end{figure}

In Fig. \ref{figure.qqs}, we can see the transition of the
chiral condensate from the first phase to the second phase. Now,
the chiral condensate in the second phase tends to decrease
abruptly with the chemical potential.
 This effect is enhanced by
temperature. 
Keeping in mind the previous section, we can set the
constant $\tilde h_1\sim 0$ if the G-MOR relation is valid.

\subsection{Isospin-number density.}

The non-vanishing components of the condensed vector current at
order $s^0$ in the loop corrections are
\begin{eqnarray}
\llang 0|\bm{V}_\mu|0\rrang_{(1,0)}&=&-f^2|\mui|s^2u_\mu{\bf e}_3,\\
\llang 0|\bm{V}_\mu|0\rrang_{(1,2)} &=& \llang 0|
-i(\tilde\pi_1\partial_\mu\tilde\pi_2-\tilde\pi_2\partial_\mu\tilde\pi_1)
\nonumber \\
&& -2|\mui||\tilde\pi|^2u_\mu+ |\mui|^3{\cal O}(s^2)|0\rrang{\bf
e}_3.
\end{eqnarray}
Like in the firs phase, the vacuum expectation value of the vector current will
be oriented as $\llang
0|\bm{V}_{\mu}|0\rrang =\llang n_I\rrang u_\mu{\bf e}_3$.
The isospin number density condensate at finite temperature and
chemical potential is then
\begin{equation}
\llang n_I\rrang=-|\mui|f^2\big\{s^2+8\alpha'J'\big\}.
\end{equation}

\begin{figure}
\includegraphics[scale=.8]{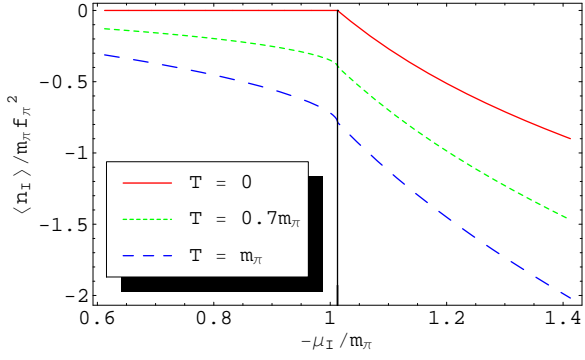}
\caption{Isospin number density as a function of the isospin
chemical potential for different values of the temperature.}
\label{figure.nis}
\end{figure}

In fig. \ref{figure.nis}, we can see the transition of the isospin
number density from the first phase to the second phase. 
Like the
chiral condensate, in the second phase $\llang n_I\rrang$ tends to
decrease abruptly with the chemical potential. 
Once again this
effect is enhanced by temperature.

\subsection{Pion condensate.}

Now, in the second phase, the pion condensate is finite and
provides us important information about the condensed phase. The
pion condensate is defined as
\begin{equation}
\llang\pi^a\rrang\equiv \llang 0|i\bar q\gamma_5\tau^aq|0\rrang
=\llang 0|J_p^a|0\rrang,
\end{equation}
and carry the same quantum numbers of the pion field. The
non-vanishing components of the pion condensates are
\begin{eqnarray}
\bm{J}_{p(1,0)} &=&-2Bf^2s\tilde{\bf e}_1,\\
\bm{J}_{p(1,2)}
&=& Bs\big[\pi_0^2+2|\pi|^2\big]\tilde{\bf e}_1,\\
\bm{J}_{p(3,0)} &=&-4B\mui^2s[l_3+l_4+{\cal O}(s^2)]\tilde{\bf
e}_1.
\end{eqnarray}
As we did for the chiral condensate, we must keep the term $s$,
since it is a global factor. We can see that the pion condensate
is oriented in the direction $\tilde{\bf e}_1$, i.e.,
$\llang\bm{\pi}\rrang=\llang\tilde\pi_1\rrang\tilde{\bf e}_1$.

Proceeding in the same way as we did with the other condensates,
the pion condensate at finite temperature and isospin chemical
potential is
\begin{equation}
\llang\tilde\pi_1\rrang =2Bf^2s\{1+\alpha'[-\fr{1}{2}\bar l_3
+2\bar l_4 -2I'_0-4I']\}.
\end{equation}
It is important to recall that the pion condensates, like the
chiral condensate, give an information about the quarks condensed
in the vacuum; in this case the pion condensate is a mixture of
$\bar u\gamma_5d$ and $\bar d\gamma_5 u$. Although related, do not
confuse the pion condensate with the number density of condensed
pions.

Note that this result is basically the same as the chiral
condensate (except the global factors $s$ and $-c$) if we neglect
the term $\tilde h_1$. Then the chiral condensate and pion
condensate satisfy the relation
\begin{equation}
s\llang\sigma\rrang +c\llang\tilde\pi_1\rrang=0. \label{sqq+cpi}
\end{equation}
We will see that this relation is valid in the limit when
$|\mui|\gg m$.

\begin{figure}
\includegraphics[scale=.8]{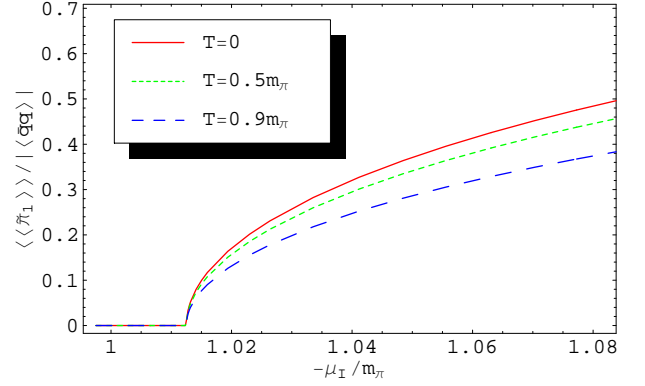}
\caption{ Pion condensate as a function of the
isospin chemical potential for different values of the
temperature. The vertical line denotes the transition point
between the two phases.} \label{figure.pis}
\end{figure}

Fig. \ref{figure.pis} shows the behavior of the pion condensate as
function of $\mui$ for different values of $T$. If we compare with
fig. \ref{figure.qqs}, we can see that the chiral condensate
diminishes with the chemical potential in a similar rate as the
pion condensate increases, according to eq. (\ref{sqq+cpi}). 
This
suggests that the quarks in the chiral condensate may mix together
to form a mixed $u$-$d$ pseudo-scalar Cooper-pairs state. Because of
the thermal bath, both chiral and pion condensates decrease as a
function of $T$, as expected, due to the increase of the $\pi^-$
mass in the thermal bath.

\subsection{Axial-isospin charge density (AICD) condensate}

The second phase supports also another condensate, the
axial-isospin charge density condensate, defined as
\begin{equation}
\llang {\cal Q}_A^a\rrang =\llang
0|\fr{1}{2}q^\dag\gamma_5\tau^aq|0\rrang = \llang 0|A_0^a|0\rrang
\end{equation}
The non-vanishing components  of the effective axial current,
according to the chiral approach, in the vacuum at the lowest
order in the $s$ expansion are
\begin{eqnarray}
\llang 0|\bm{A}_{\mu(1,0)}|0\rrang &=& f^2|\mui|cs\tilde{\bf e}_2,\\
\llang 0|\bm{A}_{\mu(1,2)}|0\rrang &=& -|\mui|cs\llang
0|\frac{i}{|\mui|}
(\tilde\pi_+\partial_\mu\tilde\pi_-
-\tilde\pi_-\partial_\mu\tilde\pi_+)\nonumber\\
&&
\quad+\pi_0^2+3|\pi|^2 +\mui^2{\cal O}(s^2)   |0\rrang\tilde{\bf e}_2
,\\
\llang 0|\bm{A}_{\mu(3,0)}|0\rrang &=& |\mui|^3cs[2l_4 +{\cal O}(s^2)].
\end{eqnarray}
Note that this axial-charge density is oriented in the $\tilde{\bf
e}_2$ direction:$ \llang 0|\bm{A}_\mu|0\rrang = \llang {\cal\tilde Q}_A^2\rrang
u_\mu\tilde{\bf e}_2$,
perpendicular to the orientation of the pion condensate and also
with respect to the Vector-current vacuum expectation value. The
three previous condensates are a basis in the isospin $SU(2)$
space. The AICD condensate is then
\begin{equation}
\llang {\cal \tilde Q}_A^2\rrang = f^2cs[1+ \alpha'(2\bar
l_4-4I'-4I'_0-8J')].
\end{equation}

\begin{figure}
\includegraphics[scale=.5]{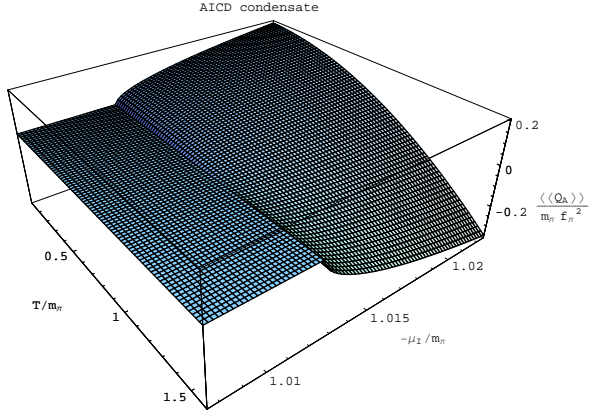}
\caption{Axial-Isospin charge density condensate, scaled with
$m_\pi f_\pi^2$, plotted in the neighborhood of the phase
transition as function of temperature and isospin chemical
potential, scaled both with $m_\pi$.} \label{figure.qas}
\end{figure}

Fig. \ref{figure.qas} shows the behavior of the AICD condensate
as function of temperature and $\mui$. Note that at a certain
critical temperature, for growing values of $\mui$, the AICD
condensate changes its sign. We would like to remark that this
object seems to be a very interesting order parameter for the
phase transition into the pion superfluid phase. Perhaps lattice
measurements could confirm this point.

\section{Second phase: $|\mui|\gg m$}

When $|\mui|\gg m$, the natural expansion parameter is
$c=\frac{m^2}{\mui^2}$. So the different objects will be expressed
as power series of the form
\begin{equation}
\llang J\rrang =\langle J\rangle \bigg\{\eta
(c)+\alpha'\sum_{n=0}\big[\sigma_n(T/|\mui|)+\sigma^{\mbox{\tiny log
}}_n\ln{c}\big]c^{n}
\bigg\} .
\end{equation}
The main difference between the previous equation and eq.
(\ref{cond.s}) is the presence of logarithms since loop
corrections give rise to terms of the form
$\lambda-\ln(\mui^2/\Lambda^2)$, which cancel with the terms
$\lambda-\ln(m^2/\Lambda^2)$ coming from the $l_i$ coupling
constants. These logarithms turn out to be extremely important for
the behavior of the condensates in this region.

\subsection{Chiral condensate}

The non-vanishing components of the chiral condensate, neglecting higher
corrections of order $c^2$, are
\begin{eqnarray}
\llang 0|J_s|0\rrang_{(1,0)}&=&-2Bf^2c,\\
\llang 0|J_s|0\rrang_{(1,2)}&=&Bc\llang
0|\pi_0^2+\tilde\pi_1^2+\tilde\pi_2^2|0\rrang,\\
\llang 0|J_s|0\rrang_{(3,0)}&=&-2B\mui^2c[l_4+2\tilde h_1+{\cal O}(c^2)],
\end{eqnarray}
keeping the global term $c$. Following the same procedure as in the previous
chapters, the chiral condensate at finite temperature and isospin chemical
potential is
\begin{multline}
\llang\sigma\rrang=-2Bf^2c\bigg\{1+\alpha'\bigg[\bar l_4+32\pi^2\tilde h_1\\
+2\ln c-4I'_0
-\fr{2}{3}\left(\fr{\pi T}{\mui}\right)^2\bigg]\bigg\}.
\label{equation.qqc}
\end{multline}

\begin{figure}
\includegraphics[scale=.75]{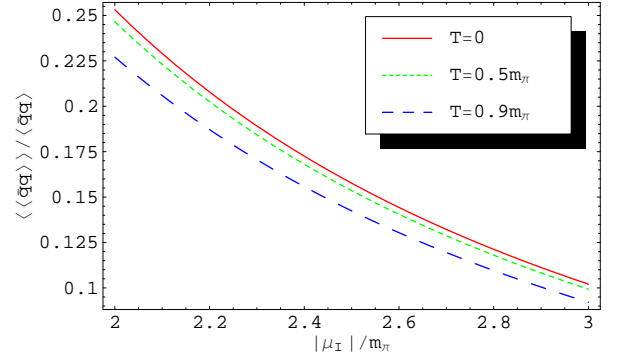}
\caption{Chiral condensate as a function of the isospin chemical
potential for different values of the temperature for high values of the
chemical potential.}
\label{figure.qqc}
\end{figure}

Fig. \ref{figure.qqc} shows the chiral condensate behavior at
finite temperature for high values of the chemical potential. It
has a decreasing behavior as function of both parameters.

\subsection{Isospin number density.}

The non-vanishing components of the vector current, within the thermal
vacuum, neglecting higher corrections of order $c^2$, are

\begin{eqnarray}
\llang 0|\bm{V}_\mu|0\rrang_{(1,0)}&=&-|\mui|f^2s^2u_\mu{\bf e}_3,\\
\llang 0|\bm{V}_\mu|0\rrang_{(1,2)}&=&|\mui|s^2\llang
0|\pi_0^2+\tilde\pi_1^2 +\mui^2{\cal O}(c^2)|0\rrang u_\mu{\bf e}_3,\nonumber\\
&&\\
\llang 0|\bm{V}_\mu|0\rrang_{(3,0)}&=&-4|\mui|^3s^2\big[l_1+l_2+{\cal O
}(c^2)\big]u_\mu{\bf e}_3.
\end{eqnarray}
The isospin number density is then
\begin{equation}
\llang n_I\rrang=-|\mui|f^2s^2\left\{1
+2\alpha'\left[\fr{1}{3}\bar l_1+\fr{2}{3}\bar l_2+2\ln c-4I'_0\right]\right\}.
\end{equation}

\begin{figure}
\includegraphics[scale=.7]{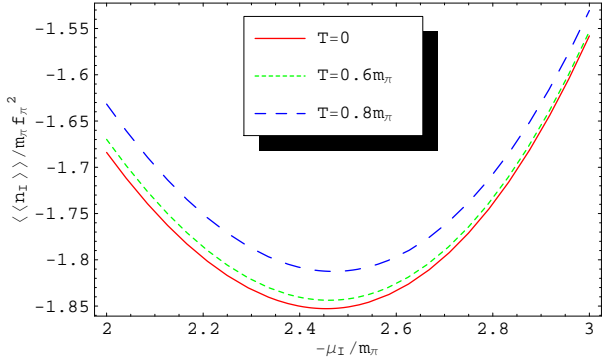}
\caption{ Isospin-number density as a function of the isospin
chemical potential at different values of the temperature in the high chemical
potential limit.}
\label{figure.nic}
\end{figure}

Fig. \ref{figure.nic} shows the isospin number density for
different temperatures and high values of the isospin chemical
potential. Due to the logarithmic term, it starts to grow again
with the chemical potential. As in the case of the $\pi^+$ mass in
this high chemical potential limit, it grows with temperature, in
contrast to the $|\mui|\sim m$ case. A crossover of the different
temperature lines must occur somewhere in the intermediate region
of the chemical potential. At this point, however, we must proceed
with care since two loop corrections could provide us with other
logarithmic terms that may change the shape of this curve. In the
previous case, close to the phase transition point, we do not have
such difficulties to deal with.

\subsection{Pion condensate}

The non-vanishing components of the pseudo-scalar current,
within the thermal
vacuum and neglecting higher corrections of order $c^2$, are
\begin{eqnarray}
\bm{J}_{p(1,0)} &=& 2Bf^2s\tilde{\bf e}_1,\\
\bm{J}_{p(1,2)} &=&
-Bs\big[\pi_0^2+\tilde\pi_1^2+\tilde\pi_2^2\big]\tilde{\bf e}_1,\\
\bm{J}_{p(3,0)} &=& 2\mui^2Bs\big[l_4+{\cal
O}(c^2)\big]\tilde{\bf e}_1.
\end{eqnarray}
The pion condensate is then
\begin{equation}
\llang\tilde\pi_1\rrang=2Bf^2s\bigg\{1+\alpha'\bigg[\bar l_4
+2\ln c-4I'_0-\left(\fr{\pi T}{\mui}\right)^2\bigg]\bigg\}.
\end{equation}

\begin{figure}
\includegraphics[scale=.7]{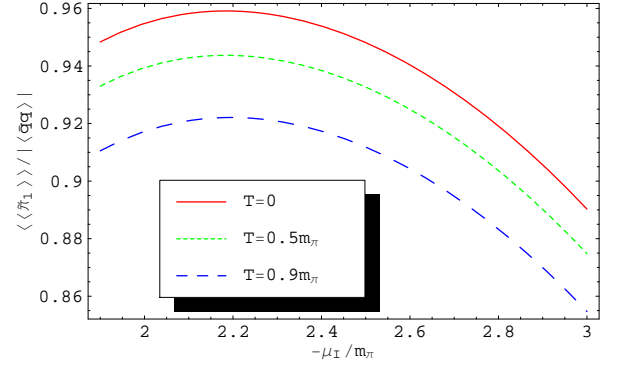}
\caption{Pion condensate as a function of the isospin chemical
potential for different temperatures in the limit of high chemical potential.}
\label{figure.pic}
\end{figure}

Fig. \ref{figure.pic} shows the pion condensate at finite
temperature and high values of the chemical potential. Like the
isospin number density, it has a deflection and starts to
decrease, due to the logarithmic terms. Again in this case, we
must be careful since as in the isospin number density evolution,
new important logarithmic terms, from two-loop calculations, could
also change the bending as well as the general shape of our
one-loop result.
 Comparing with equation
(\ref{equation.qqc}), if we neglect the term $\tilde h_1$, the
pion condensate and the chiral condensate
 follow the relation $s\llang\sigma\rrang
+c\llang\tilde\pi_1\rrang=0$, as was the case in the region $|\mui|\gtrsim m$.

\subsection{AICD condensate}

The axial-vector current components, neglecting higher terms in
the expansion in terms of $c^n$ are
\begin{eqnarray}
\bm{A}_{\mu(1,0)} &=& f^2|\mui|csu_\mu\tilde{\bf e}_2,\\
\bm{A}_{\mu(1,2)} &=& |\mui|cs\bigg[ \frac{1}{|\mui|c}
(\tilde\pi_1\partial_\mu\tilde\pi_2
-\tilde\pi_2\partial_\mu\tilde\pi_1)\nonumber\\
&&
\hspace{2.2cm}-\pi_0^2-2\tilde\pi_1^2-\tilde\pi_2^2 \bigg]\tilde{\bf e}_2\label{Amu(1,2).c} \\
\bm{A}_{\mu(3,0)} &=& |\mui|^3cs2\big[l_1+l_2 +{\cal
O}(c^2)\big]u_\mu\tilde{\bf e}_2,
\end{eqnarray}
where we keep the term $cs$ as a global factor. In this case we
use also the propagator $D_{DJ}(p)_{12}$ which is of order $c$.
The factor $1/c$ in eq. (\ref{Amu(1,2).c}) cancels the other term
$c$ of the propagator, so our result is still of ${\cal O}(c^0)$.
As int the case $|\mui|\gtrsim m$, the AICD condensate will be
oriented in the $u_\mu\tilde{\bf e}_3$ direction. The AICD condensate is then
\begin{multline}
\llang {\cal \tilde Q}_A^2\rrang =
f^2cs\bigg\{1+\alpha'\bigg[\fr{2}{3}\bar l_1+\fr{4}{3}\bar l_2 +2 \ln c
-\fr{1}{2}\\
-12I'_0+I'_1
 -\fr{2}{3}\left(\fr{\pi
T}{\mui}\right)^2-\fr{32}{15}\left(\fr{\pi T}{\mui}\right)^4\bigg]
\bigg\}.
\end{multline}

\begin{figure}
\includegraphics[scale=.5]{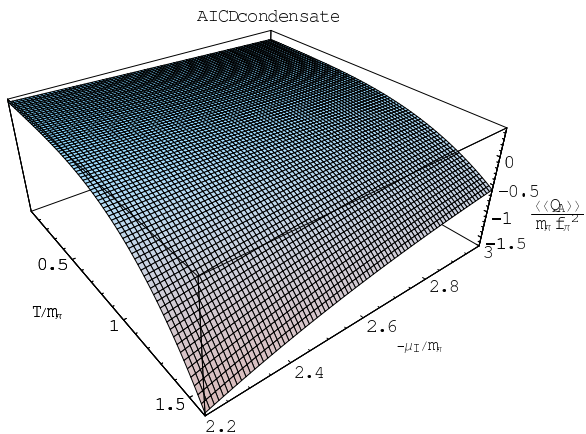}
\caption{Axial-Isospin charge density condensate, scaled with
$m_\pi f_\pi^2$, plotted for high values of $\mui$, as function of
temperature and isospin chemical potential, scaled both with
$m_\pi$.} \label{figure.qac}
\end{figure}

Fig. \ref{figure.qac} shows the behavior of the AICD condensate
for high values of $\mui$. Due to the logarithmic term, the
condensate for high $T$ values starts to grow again as function of
$\mui$.

\section{Final comments and summary}

It is interesting to remark that our pion condensate behaves as
the diquark condensate, according to the lattice QCD analysis, as
function of temperature and isospin chemical potential
\cite{Kogut:2002tm,Kogut:2002zg}. This condensate starts to rise
as soon as we go into the second phase, close to the phase
transition point \cite{Kogut:2001if,Kogut:2002cm}. However, for
bigger values of $\mui$ it decreases, in agreement with our
analytical results which show a decreasing behavior as $\sim\ln
1/\mui^2$. The lattice analysis has confirmed also the relation
between the chiral condensate and the pion condensate according to
eq. (\ref{sqq+cpi}). The chiral condensate and the isospin number
density also agree with the lattice results.

We would like to emphasize that the G-MOR  relation (at least in
the first phase) remains valid if we write it in terms of the
average values of the pion masses and decay constants. Also the
relation $s\llang\sigma\rrang +c\llang\tilde\pi_1\rrang=0$
is confirmed lattice results \cite{Kogut:2002cm}.

Finally, we think that
the realization of condensates as a basis
in the isospin space in the second phase ($\llang \bm{A}_0\rrang$,
$\llang \bm{V}_0\rrang$, $\llang \bm{J}_p\rrang$ are all
perpendicular) is a very interesting phenomenon. One-loop thermal
corrections do not destroy this property, which remains valid when
$|\mui|\gtrsim m$ as well as when $|\mui|\gg m$.


\subsection*{Acknowledgements}
The work of  M.L. and C.V. have been
supported
 by Fondecyt (Chile)
under grant No.1010976.

\bigskip
\mbox{}
\bigskip

\section*{APPENDIX}

The functions involved in the radiative corrections in the first phase are
\begin{eqnarray*}
\alpha  \!&=& \!(m/4\pi f)^2\\
I   &=& \int_1^\infty \!\!\! dx\sqrt{x^2-1}
        [n_B(mx-|\mui|)+n_B(mx+|\mui|)]\\
J   \!&=&\! \int_1^\infty \!\!\! dxx\sqrt{x^2-1}
        [n_B(mx-|\mui|)-n_B(mx+|\mui|)]\\
I_n \!&=&\!  \int_1^\infty \!\!\! dxx^{2n}\sqrt{x^2-1}~2n_B(mx)
\end{eqnarray*}
being $\alpha$ the perturbative parameter. These integrals do not depend on
the chemical potential sign, and grow with both, temperature and chemical
potential.

In the case of the second phase, we
denote the same functions with a prime: $\alpha'$, $I'$, $J'$, $I'_n$ which are
the same functions, but with $|\mui|$ instead of $m$.
\begin{eqnarray*}
\alpha' \!&=&\!\! (\mui/4\pi f)^2\\
I'   \!&=&\!\! \int_1^\infty \!\!\! dx\sqrt{x^2-1}
\big[n_B\big(|\mui|(x\!-\!1)\big)+n_B\big(|\mui|(x\!+\!1)\big)\big]\\
J'   \!&=&\!\! \int_1^\infty \!\!\! dxx\sqrt{x^2-1}
\big[n_B\big(|\mui|(x\!-\!1)\big)-n_B\big(|\mui|(x\!+\!1)\big)\big]\\
I'_n \! &=&\! \! \int_1^\infty \!\!\! dxx^{2n}\sqrt{x^2-1}~2n_B(|\mui|x).
\end{eqnarray*}
These integrals are also growing functions of the temperature but
decrease with the chemical potential.



\vfill
\bibliography{bib}

\end{document}